\documentclass[10pt,conference]{IEEEtran}
\IEEEoverridecommandlockouts

\usepackage{cite}
\usepackage{amsmath,amssymb,amsfonts}
\usepackage{algorithmic}
\usepackage{graphicx}
\usepackage{textcomp}
\usepackage{xcolor}
\usepackage{enumerate}
\usepackage{url}
\usepackage{todonotes}
\usepackage{comment}
\usepackage{booktabs}
\usepackage{hyperref}

\def\BibTeX{{\rm B\kern-.05em{\sc i\kern-.025em b}\kern-.08em
    T\kern-.1667em\lower.7ex\hbox{E}\kern-.125emX}}

\begin{document}

\title{Power Assumptions Matter: Evaluating End-user Laptop Energy Models for Sustainability Reporting of Browser-Based Web Services
\thanks{This work was supported by the Independent Research Fund Denmark Project no. 2102-00281B.}
}

\author{\IEEEauthorblockN{1\textsuperscript{st} Maja H.\ Kirkeby}
\IEEEauthorblockA{\textit{Computer Science, Dep.\ of People and Technology} \\
\textit{Roskilde University}\\
Roskilde, Denmark \\
kirkebym@acm.org 0000-0003-0033-2438}
\and
\IEEEauthorblockN{2\textsuperscript{nd} Timmie Lagermann}
\IEEEauthorblockA{\textit{Computer Science, Dep.\ of People and Technology} \\
\textit{Roskilde University}\\
Roskilde, Denmark \\
0009-0002-4728-0305}
}

\maketitle

\begin{abstract}
Sustainability reporting for web-based services often relies on simplified end-user energy models that assume constant laptop power during browser interactions. Energy models such as Digst and DIMPACT apply fixed power values (15--22~W), yet the validity of this approach for realistic browsing remains underexplored.

We empirically evaluate constant-power assumptions in a controlled user study where ten participants repeatedly complete eight representative user flows across shopping, booking, navigation, and news services on four laptop platforms, while device energy is measured. Typical power is 9--13~W, substantially below current reporting standards, implying systematic overestimation. Moreover, the error scales proportionally with task duration, indicating systematic bias rather than random noise.

Comparing progressively refined constant-power models, we find that category-specific parameters improve accuracy more than hardware-only parameters and approach flow-specific performance. The best fit is obtained by combining category (or flow) with hardware, while category-level models retain most of the benefit with fewer parameters, making them a practical upgrade for sustainability reporting.

\end{abstract}

\begin{IEEEkeywords}
Energy model, Web services, Laptop, Empirical Evaluation, Sustainability Reporting
\end{IEEEkeywords}

\section{Introduction}




Digital sustainability reporting has become central to the evaluation of web-based services. Governments and organizations increasingly depend on simplified models to estimate energy use and greenhouse gas emissions. Two such models are the Danish Agency of Digital Government's Digst framework~\cite{digst} and the UK-based DIMPACT framework~\cite{dimpact,DIMPACTtool2022} based on~\cite{Preist2019}. However, despite their growing role, the empirical validation of these models is limited. Without systematic assessment, it remains unclear how accurately they capture the energy usage of real-world web interactions.

This paper addresses this gap by presenting an experiment that directly compares measured energy consumption against model-based estimates. 
Following the recommendation by Guldner et al.~\cite{Guldner2024} to evaluate software energy efficiency under representative user scenarios, and inspired by the use of online-media categories by Schien et al.~\cite{Schien2013}, we focus on primarily non-streaming web services: shopping, booking, navigation, and news.
While categories such as audio/video streaming and generative AI are highly relevant to digital sustainability, their energy use depends on complex scaling parameters, e.g., ~\cite{Suski2020,Luccioni2024}, that require separate, parameter-specific studies. We therefore delimit our study to web services with more stable and comparable interaction patterns.

Existing sustainability reporting frameworks for web-based services, such as 
Digst and DIMPACT, estimate end-user device energy use through constant-power 
assumptions. 
While these models enable comparability across services, their empirical 
accuracy and the factors influencing their validity remain largely unexplored. 
To address this gap, we investigate two research questions:
\begin{description}
\item[RQ1:] How accurately do constant-power assumptions in current reporting frameworks approximate measured end-user device energy use during realistic browser-based user flows?
\item[RQ2:] How does accuracy change when constant-power models are parameterized by device, service category, and user flow, and which factors contribute most to improved precision?
\end{description}
%
%
%
To answer these, we make the following contributions:
(1)~\textit{Protocol \& dataset:} a controlled method (and dataset) for measuring end-user device energy during realistic browser flows across services and laptops.
(2)~\textit{Duration-proportional bias:} empirical evidence that constant-power assumptions induce systematic error that scales with interaction duration.
(3)~\textit{Reporting refinement:} a quantified trade-off analysis for device/category/flow parameterization, identifying category-level calibration as a low-cost accuracy gain.

The paper is organized as follows. Section~II reviews related work on ICT sustainability reporting and web-energy measurement. Section~III describes the study design, measurement setup, and model evaluation procedure. 
Section~IV presents results for RQ1 and RQ2, and Section~V discusses threats to validity. Section~VI interprets implications for sustainability reporting models, and Section~VII concludes with future work.

\section{Background and Related Work}
This section situates the study within existing research on the estimation 
and reporting of Information and Communication Technology (ICT) related energy use. 
We will focus on how they deal with use-phase energy consumption for end-user devices.
We review Sector-level and Service-level models for reporting, software-level energy-efficiency measurement frameworks, and the web-specific energy efficiency. 
We then outline software- and measurement-level approaches that aim to 
quantify energy use more precisely at the program or interaction level. 
Together, these works frame the methodological gap that this 
paper addresses: linking reproducible, software-level energy measurements 
to the simplified models used in large-scale sustainability reporting.

\subsection{Sector-Level Reporting Models}
Global and sector-level ICT footprint models establish the broader context for digital sustainability research. Early studies, e.g.,~\cite{Andrae2015,Malmodin2018,Belkhir2018}, raised awareness of the growing energy challenges, while Freitag et al.~\cite{Freitag2021} analyzed the methodological assumptions. Later studies, such as Malmodin et al.~\cite{malmodin2024}, provide comprehensive evaluation of the ICT sector’s total energy use and greenhouse gas emissions.  

These models typically employ top-down allocation and averaged intensity factors to distribute total sector energy across subdomains. As a result, they inherently contain uncertainties, and Malmodin et al.~\cite{malmodin2024} specifically acknowledge that \textit{``the number of user devices in operation and their average electricity consumption remain a more uncertain part in line with previous studies''}. This uncertainty is echoed by Furberg et al.~\cite{Furberg2026}, who note that \textit{``to estimate the electricity use of devices, data on actual device usage patterns (e.g., operational service time per year) are required. In practice, studies on the ICT sector’s direct climate impact usually do not collect device usage data but rely on secondary data from various sources''}, e.g., averages over multiple sources~\cite{malmodin2024}.

In relation to our study, these studies provide an essential macro-level baseline for ICT sustainability. While uncertainty is inherent to top-down footprinting, it is notable that end-user devices' electricity consumption and usage patterns are emphasized as a particularly uncertain component~\cite{malmodin2024,Furberg2026}. These models relate to service-level models, which operationalize this approach in actionable estimates at the granularity of specific services and user interactions

\subsection{Service-Level Reporting Models}
Service-level models aim to make the environmental assessment of digital services practical and comparable. Frameworks such as the Danish Agency of Digital Government\footnote{\url{https://en.digst.dk/}}'s Digst model~\cite{digst} and the United Kingdom-based DIMPACT tool~\cite{dimpact,DIMPACTtool2022} operationalize simplified estimation methods that express energy use as the product of assumed constant device power and activity duration~($P \times t$). 

These models allocate a share of data center and network energy to each service and are increasingly used for organizational sustainability reporting. The approach aligns with academic efforts such as Preist et al.~\cite{Preist2019}, who analyzed how interface design in YouTube affects energy demand, and with the Green Software Foundation’s Software Carbon Intensity (SCI) specification, which formalizes functional units and allocation principles for shared infrastructure.

Although these frameworks facilitate transparency and comparability, they rely on several simplifying assumptions. Notably, they carry over the sector-level practice of using averaged end-user device parameters by assuming constant device power, alongside assumptions of uniform user behavior and linear relationships between time, data volume, and energy use.

The precision and accuracy of these models have not yet been evaluated against measured data from real user interactions. 
Our study directly addresses this gap by empirically testing the validity of the $P \times t$ assumption across realistic user flows and hardware configurations. By comparing measured energy data with estimates derived from Digst- and DIMPACT-style models, we provide an evidence-based assessment of their accuracy and identify conditions under which these models deviate from actual energy consumption. 

\subsection{Software-Level Efficiency}
An orthogonal and very active stream of research in software-level energy efficiency has produced a range of tools and methodologies for analyzing and improving the energy behavior of programs. While these cannot be used for reporting since they do not provide the absolute energy consumed, but instead provides relative energy consumed and can therefore provide insights to which programs consume more energy. 
A series of tools EACOF~\cite{Field2014}, PowerAPI~\cite{Fieni2024}, Keppler~\cite{Amaral2023} --a comparison of these and other related tools are given by Jay et al.~\cite{jay2023}-- enabled applications to access power data at runtime, fostering awareness of energy use during development. These tools uses Intel's RAPL~\cite{Intel2020RAPL,Intel64IA32Vol3B} which estimates the energy consumed with relative precision and we cannot use this approach for reporting. 

\subsection{Measurement Frameworks}
More recently the research has also focused on aligning the methodology used when measuring, e.g., Mancebo et al.~\cite{Mancebo2021} and Guldner et al.~\cite{Guldner2024}. Mancebo et al. have focused on the procedure and documentation of the procedure and we follow the recommendations and provide the full information for our experiment in an open repository~\texttt{EUD-energy-model}\footnote{See \url{https://github.com/SustainableSoftware/EUD-energy-model.git}}. 
Guldner et al. propose a standardized process for measuring software resource efficiency under representative user scenarios. Their rationale is that meaningful and reproducible results require measurements grounded in realistic user interactions rather than synthetic benchmarks. This principle directly aligns with our methodological choice to evaluate energy consumption through end-to-end user flows across web services.




\subsection{Web-Specific Energy Studies}
Prior work has examined web-service energy use through both modeling and measurement. Schien et al.~\cite{Schien2013} modeled online media use across text-based, interactive, and streaming content and found that end-user devices—particularly laptops—dominate overall energy consumption, motivating categorization by interaction type and data intensity, which we also adopt.
%

Subsequent studies examined factors driving variability. Suski et al.~\cite{Suski2020} showed that user choices (e.g., device and video resolution) affect streaming emissions. 
Methodologically, Katsenou et al.~\cite{Katsenou2024} compared external and RAPL-based energy measurements for video compression, and Herglotz et al.~\cite{Herglotz2025} developed synchronized measurement procedures for video communication, 
showing that RAPL correlates with external measurements but remains inaccurate.
In our data, one news flow included embedded video playback, enabling comparison with these controlled observations.

Collectively, prior work demonstrates that web-service energy use is heterogeneous and context dependent. However, few of these studies examine whether the simplified assumptions used in service-level reporting models can capture such variability. Our work addresses this gap by providing measured, category-based data to test the validity of those assumptions under realistic web interactions.

\section{Methodology}
 This section describes the essential parts of our experimental design, data collection, and model evaluation procedures used to address the research questions. 

All anonymized user-flow descriptions, hardware, physical setup, scripts, software versions, datasets, and analysis scripts are 
available in the open repository~\texttt{EUD-energy-model}\footnote{See \url{https://github.com/SustainableSoftware/EUD-energy-model.git}}. 
%

\begin{figure}[t]
    \centering
    \includegraphics[width=0.9\linewidth]{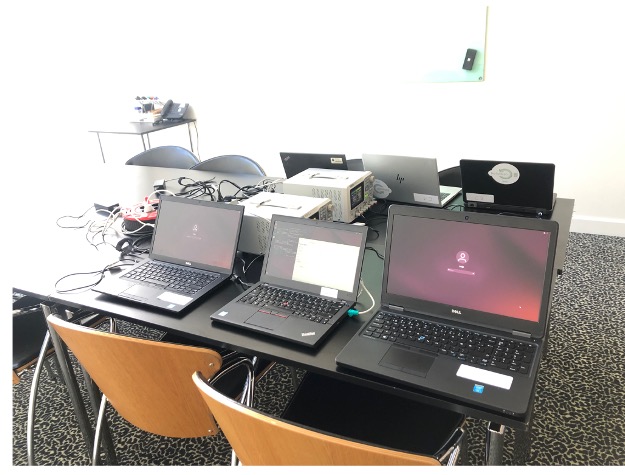}
    \caption{The setup for the experiment with 4 Laptops, two Siglents and two data-collecting computers with user-controlled synchronization software.}
    \label{fig:setup}
\end{figure}

\subsection{Websites and User Flows}
We selected eight typical websites in four categories:

\begin{enumerate}[{C}1:]
    \item Shopping: ebay.fr, amazon.fr;
    \item Booking: hotels.com, booking.com;
    \item Navigation: google.com/maps, bing.com/maps; and 
    \item News: dr.dk, tv2.dk
\end{enumerate}

For each category, we defined realistic user flows that mimic everyday
use—searching for a product, reserving a hotel room, planning a
route, or reading a news article—so that equivalent functionality was
tested across services.  Streaming and other parameter-dependent
services were deliberately excluded to ensure comparability of
interaction length and structure.


\subsection{Devices}
Experiments were performed on four Intel laptops representing different
hardware generations and OEM configurations:
\begin{enumerate}[{L}1:]
\item Dell~Latitude~7490 (8th~gen, UHD~620),
\item HP~EliteBook~830 G6 (8th~gen, alternate firmware),
\item Dell~Latitude~E5250 (5th~gen, compact chassis), and
\item Dell~Latitude~E5550 (5th~gen, larger chassis).
\end{enumerate}
This selection captures processor-generation and design-level
power variation.

\subsection{Participants and Experimental Design}
Ten participants completed the experiment; one participant (P10) 
contributed only two runs and was excluded from the mixed-effects 
analyses to ensure stable estimation of random effects. 
The remaining nine participants provided between eight and twenty-four 
valid runs each. 
The design crossed eight user flows with four laptops, and every 
flow–computer combination was executed by at least four and up to seven 
distinct participants, ensuring balanced coverage, see \texttt{EUD-energy-model}
\footnote{\url{https://github.com/SustainableSoftware/EUD-energy-model.git}}. 
To counter potential learning or fatigue effects while maintaining full 
coverage, the order of user flows and the allocation of laptops were 
randomized under constraints that ensured each flow–device combination 
was represented across participants. 
Each flow was repeated three times per device, yielding 
a total of \(164\) valid observations.

\subsection{Experimental Setup and Measurements}
All laptops ran identical operating-system images (Ubuntu Desktop 24.04.3 LTS) with default power
settings and brightness. The same Chromium browser version was used in incognito mode, and caches were cleared before start.
Each measurement recorded device power~(W) via a Siglent SPD3303X-E programmable power supply and task
duration~(s), allowing energy to be computed as integrated power
over time.  The external power-supply measurements was
user-synchronized via task start and stop events to ensure
precise alignment between user interaction and power logging. The setup can be seen in Figure~\ref{fig:setup}.
Key sources of potential variance included machine architecture,
user execution behavior, and learning effects, which were addressed
through randomization and replication. Potential influences of these factors and the sample size are discussed in Section~\ref{sec:validity}.

\subsection{Model Evaluation}
To evaluate how accurate and precise sector-level reporting models represent actual 
end-user device energy, we implemented both the Digst and DIMPACT 
formulations. 
In both frameworks, energy use is estimated as constant device power 
multiplied by interaction duration (\(E_\mathrm{model} = P_\mathrm{avg} \times t\)). 

While Digst in general uses~\cite[Table~1 ]{digst} a composite averaging over laptops (22~W), tower desktops (87~W), all-in-one PCs (81~W), and monitors (31~W) as the constant device power\footnote{The workload adjusted average power ratings by Urban et~al.~\cite{urban2017}.} 
we have extracted the laptop power of 22~W and use this as comparison for our experiment.
In comparison, DIMPACT have chosen the lower constant 15~W for 
web-browsing scenarios~\cite{dimpact,DIMPACTtool2022}\footnote{The power ratings by Urban et~al.~\cite{urban2017} without workload adjustment.}. 
We will use these constant device powers (22W and 15W) when evaluating energy models.

\section{Results}
\begin{figure}[b]
  \centering
  \includegraphics[width=0.9\linewidth]{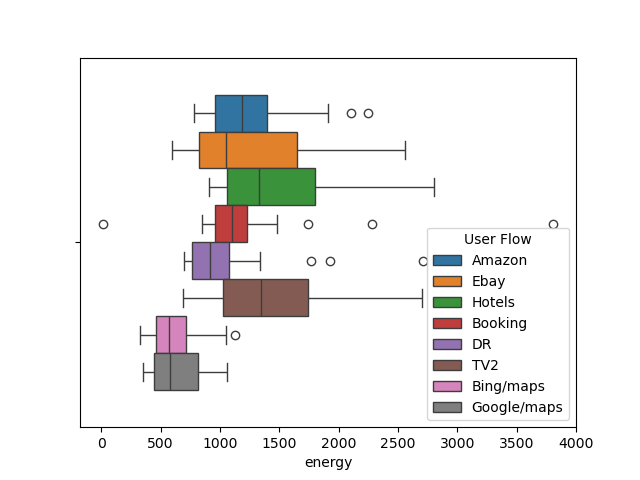}
  \caption{Distribution of measured total energy across user flows. 
  Energy varies substantially between flows, with navigation and news showing the highest medians 
  Boxes represent interquartile ranges; whiskers denote 1.5$\times$IQR.}
  \label{fig:userflow_energy}
\end{figure}

Figure~\ref{fig:userflow_energy} summarizes the measured energy used by the laptops while participants completed each user flow. Energy use varied most for shopping flows (Amazon and Ebay), whereas navigation flows (map-based tasks) had a comparatively consistent low usage.

Our results show (RQ1) systematic, duration-proportional overestimation in the Digst and DIMPACT constant-power models for realistic browser flows; and (RQ2) that adding service-category and device parameters substantially improves precision, with category-level calibration retaining most of the benefit at low complexity.

\begin{figure}[t]
    \centering
    \includegraphics[width=0.9\linewidth]{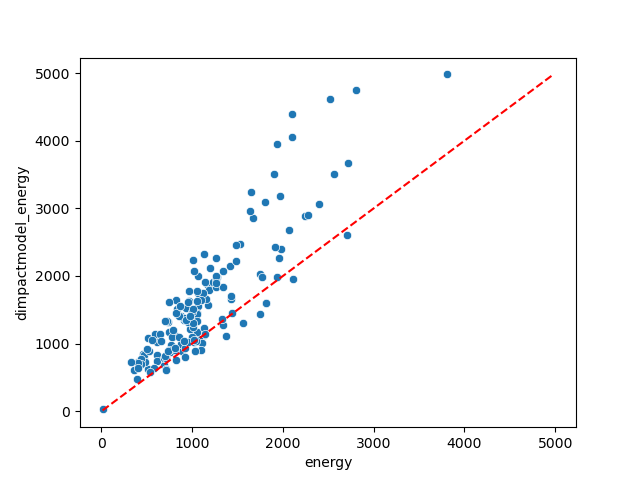}
    \caption{Measured (x-axis) versus DIMPACT's modeled energy consumption (y-axis). The red dashed line indicates perfect prediction (\(y = x\)). Many points lie above the perfect prediction; DIMPACT mostly overestimates the energy consumption.}
    \label{fig:dimpact_vs_measured}
\end{figure}

\subsection{Model Accuracy and Error Growth (RQ1)}
\label{sec:rq1}
To assess how accurately the national reporting models reflect measured device energy, 
we compared the estimated energy consumption from the DIMPACT and {Digst} 
frameworks against empirical measurements from controlled web-service user flows. 
Both models implement a simplified \(P_\mathrm{avg} \times t\) formulation, where energy use is 
approximated as constant device power multiplied by task duration. In the Digst framework, this constant is set to 22~W based on 
national reference data, whereas DIMPACT applies a lower value 
of 15~W for web-browsing scenarios. 
These fixed parameters provide a straightforward baseline for 
reporting but inherently assume uniform device behavior across 
users and workloads.

\begin{figure}[t]
    \centering
    \includegraphics[width=0.9\linewidth]{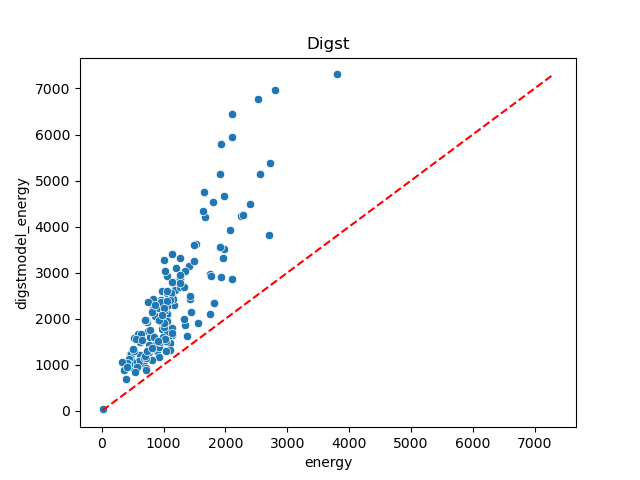}
    \caption{Measured (x-axis) versus Digst's modeled energy consumption (y-axis). The red dashed line indicates perfect prediction (\(y = x\)). All points lie above the perfect prediction; Digst overestimates the energy consumption.
    }
    \label{fig:Digst_vs_measured}
\end{figure}
\subsubsection{Model Accuracy}
Figure~\ref{fig:dimpact_vs_measured} compares measured energy with DIMPACT’s model estimates. 
The values are positively correlated, but the model systematically overestimates actual energy 
use for higher-energy tasks. Across all flows, the {mean percentage error} was 
\(26\,\%\), the {mean absolute percentage error (MAPE)} was \(28.55\,\%\), and the 
{root-mean-square error (RMSE)} was 
\(653\,\text{J}\). 

The corresponding Digst comparison (Figure~\ref{fig:Digst_vs_measured}) revealed a stronger 
overestimation bias: all observations lie above the perfect-prediction line, yielding {mean percentage error} of \(49.54\,\%\), {MAPE} of 
\(49.54\,\%\), and 
{RMSE} of \(1457\,\text{J}\). 

While both models show moderate agreement with measured values, the deviation from 
the perfect prediction line confirms that the constant-power assumption does not hold for 
realistic browser interactions.
The $R^2$-values are negative (Digst: $-5.6$, DIMPACT: $-0.3$) meaning that these predictions perform worse than always using the mean measured energy.
\begin{figure}[t]
    \centering
    \includegraphics[width=0.9\linewidth]{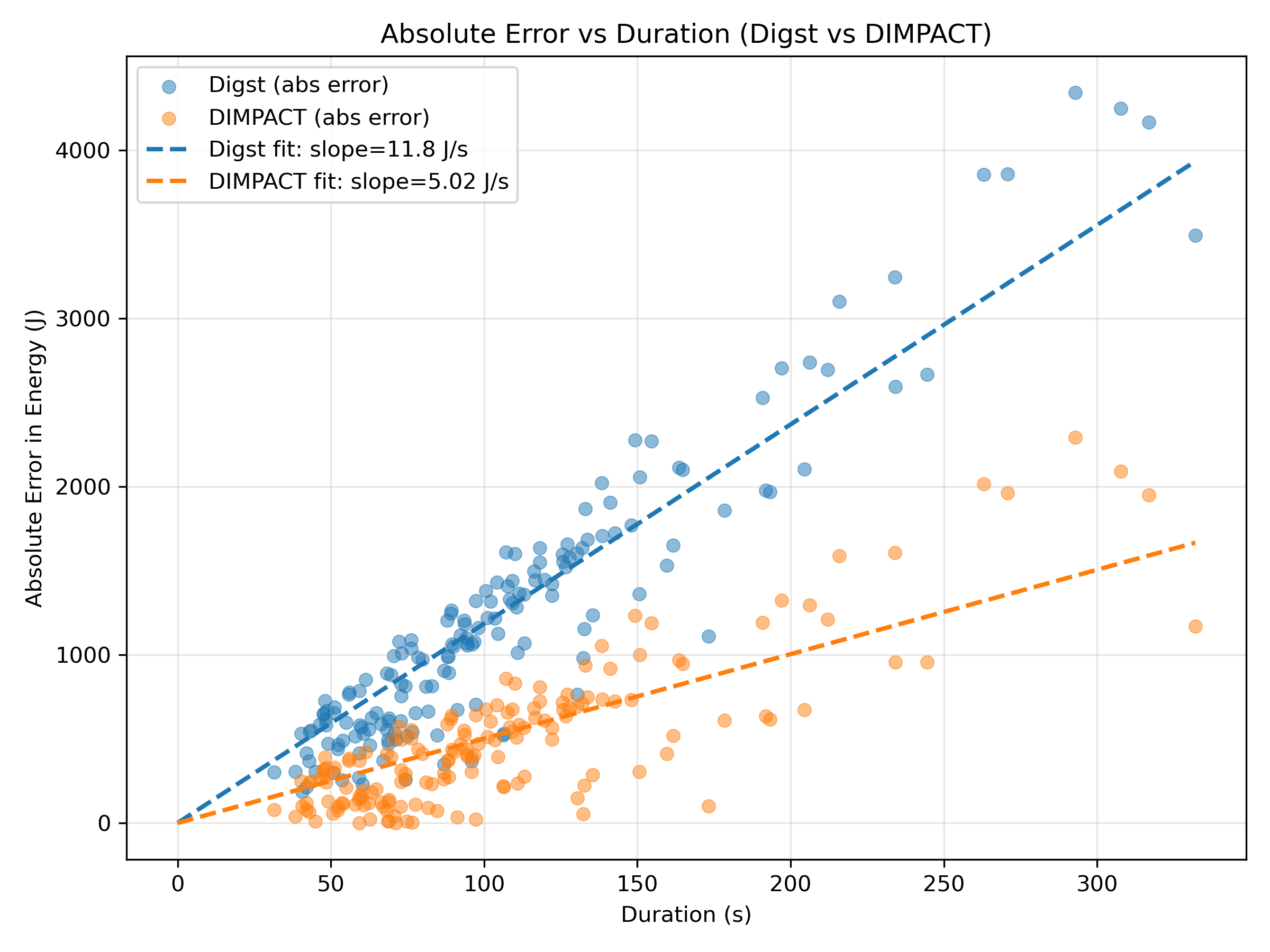}
    \caption{Absolute error in predicted energy (y-axes) versus task duration (x-axes) for Digst (blue) and DIMPACT (yellow). In both cases, the fitted through-origin model shows duration-proportional overestimation, with slopes $ 5.0\,\text{J/s}$ for DIMPACT and $11.8\,\text{J/s}$ for Digst.}
    \label{fig:dimpact_digst_error_duration}
\end{figure}
\subsubsection{Systematic Growth of Error with Duration.}
%
To assess whether error accumulates with task length, we regressed absolute energy error on task duration (Figure~\ref{fig:dimpact_digst_error_duration}). DIMPACT (yellow markers and dashed line) shows a strong duration effect ($R^{2}=0.86$, $F(1,164)=983.5$, $p<0.001$) with slope $\beta=5.0\pm0.16\,\text{J/s}$, while Digst (blue markers and dashed line) is stronger ($R^{2}=0.96$, $F(1,164)=3905$, $p<0.001$) with $\beta=11.8\pm0.19\,\text{J/s}$. In conclusion, deviations increase approximately linearly with interaction duration; per-second miscalibrations in $P_{\mathrm{avg}}!\times t$ therefore compound into systematic overestimation for longer browsing sessions.


\subsection{Factor Effects on Model Deviations (RQ2)} \label{sec:rq2}
To determine which factors explain the deviations observed in RQ1, 
we incrementally extended the constant power model with parameters 
representing hardware and web-service variability. 
All models were fitted with intercepts in (0,0) to reflect the nature of  energy usage for the formula $E = t \times P$: zero energy is used in tasks of zero seconds duration. 
Table~\ref{tab:model_comparison} summarizes the progression in explanatory power, were a (0,0) intercept is enforced by a "-1" in the formula and  "C(criteria)" indicates the partitioning criteria for the data set before training.

\begin{table}[b]
\centering
\caption{Progression of OLS models explaining measured energy (\(E\)) from duration (\(t\)) 
and contextual factors (with intercept = 0).}
\label{tab:model_comparison}
\begin{tabular}{l l c}
\toprule
\textbf{Model} & \textbf{Formula} & \textbf{\(R^{2}\)} \\
\midrule
Global best-fit & \(E \sim t - 1\) & 0.946 \\
+ Hardware (laptop) & \(E \sim t\!:\!C(\text{laptop}) - 1\) & 0.960 \\
+ Category & \(E \sim t\!:\!C(\text{category}) - 1\) & 0.973 \\
+ Flow & \(E \sim t\!:\!C(\text{flow}) - 1\) & 0.978 \\
Category $\times$ Hardware & \(E \sim t\!:\!C(\text{category})\!:\!C(\text{laptop}) - 1\) & 0.989 \\
Flow $\times$ Hardware & \(E \sim t\!:\!C(\text{flow})\!:\!C(\text{laptop}) - 1\) & 0.994 \\
\bottomrule
\end{tabular}
\end{table}

\subsubsection{Baseline}
The global model (\(E\!=\!10.15\times t\)) achieved \(R^{2} = 0.946\), corresponding to an average power of {10.15\,W}.

\subsubsection{Individual Effects}
\begin{figure}[t]
  \centering
  \includegraphics[width=0.75\linewidth]{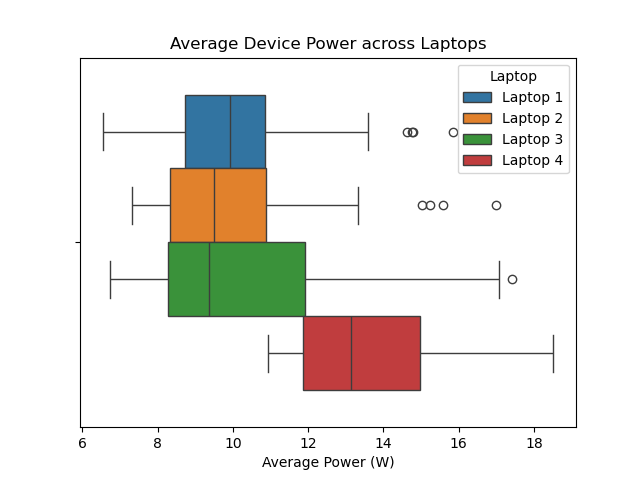}
  \caption{Distribution of average device power across laptops (L1--L4). 
  Hardware-specific differences align with the coefficients in the 
  duration-by-computer model (\(R^{2} = 0.960\)), 
  ranging from roughly 9~W for the most efficient to 12~W for the least 
  efficient device.}
  \label{fig:laptop_poweravg}
\end{figure}

\begin{figure}[t]
  \centering
  \includegraphics[width=0.9\linewidth]{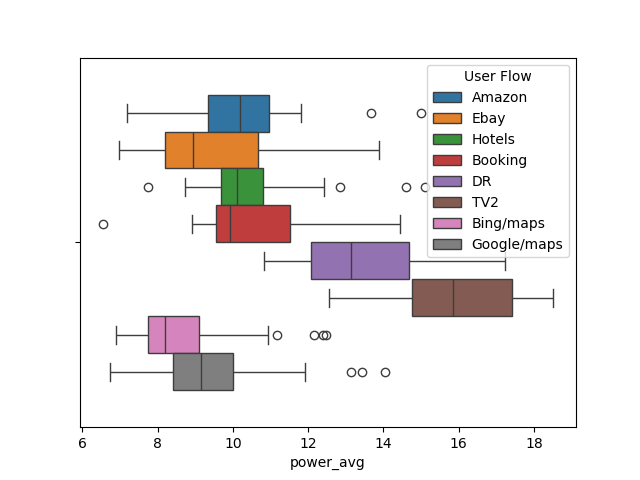}
  \caption{Distribution of average device power across user flows. 
  Power intensity varies by task type, confirming that flow-specific 
  coefficients capture systematic rather than random variance.}
  \label{fig:userflow_power}
\end{figure}

Introducing computer-specific coefficients increased the fit to \(R^{2} = 0.960\), 
revealing systematic power-model differences across devices (9.17–12.43\,W). 
These differences align with the measured power across devices shown in Figure~\ref{fig:laptop_poweravg}. 
Such hardware variation reflects both CPU generation and OEM design, and accounts for a substantial share of the total variance in energy use.
%
Adding category-specific power factors improved the model to \(R^{2} = 0.973\), 
with typical fitted powers ranging from 8.5\,W for navigation to 14\,W for news tasks. 
%
Using flow-level distinction further increased \(R^{2}\) to~0.978. Corresponding differences in average device power 
(Figure~\ref{fig:userflow_power}) confirm that these effects stem from 
task intensity rather than duration alone. 


\subsubsection{Interaction Effects.}
 When hardware and category were combined (\(C(\text{category})\!:\!C(\text{laptop})\)) obtained \(R^{2} = 0.989\). 
and the full interaction model across 
flow and computer reached \(R^{2} = 0.994\). 
These results show that nearly all residual variance stems from identifiable and differences in hardware and interaction type, rather than random error.


\section{Threats to Validity}
\label{sec:validity}
We discuss threats to validity related to participant behavior and execution
order, the operationalization of services and flows, and the generalizability
of the evaluated workload and hardware scope.

\subsection{Internal validity: order effects and participant behavior}
\begin{figure}[t]
  \centering
  \includegraphics[width=0.8\linewidth]{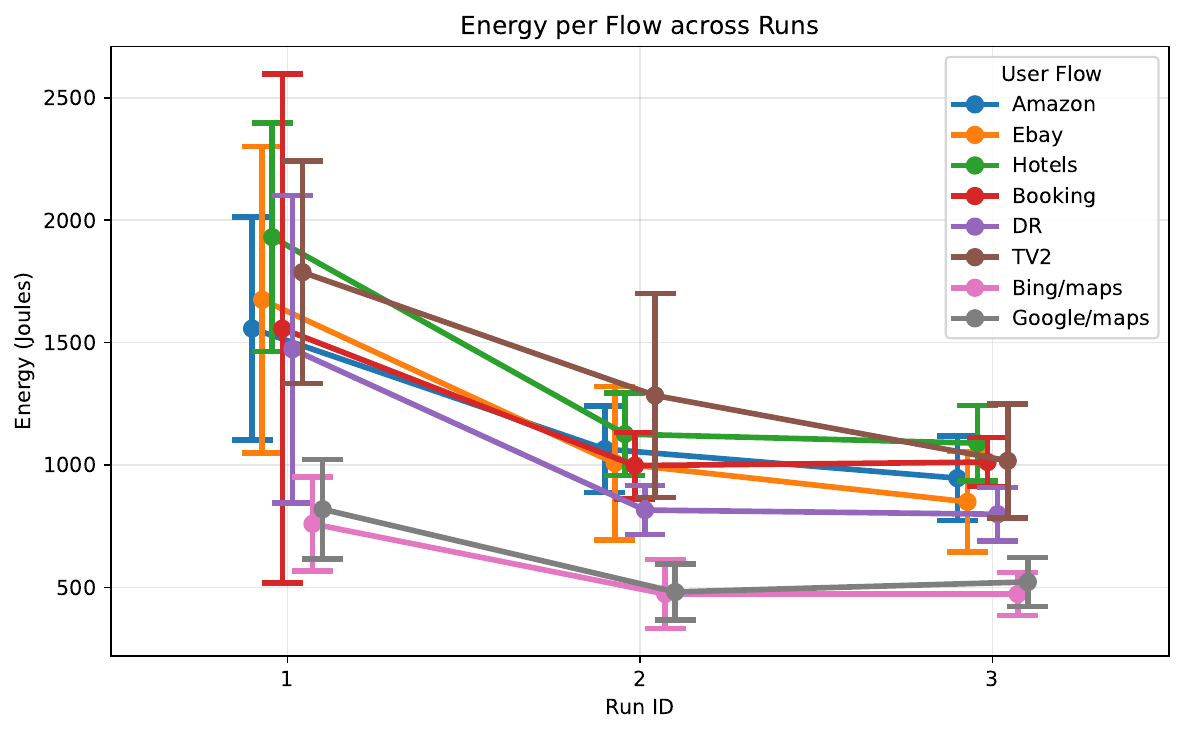}
  \caption{Average measured energy across repeated runs (1–3). 
  Energy decreased from Run~1 to Run~2 and remained stable thereafter, 
  indicating a short learning effect without fatigue. 
  Error bars represent standard deviations across participants.}
  \label{fig:run_order_effects}
\end{figure}

Repeated participation may change execution efficiency and thereby measured energy consumption. To mitigate order effects, we randomized both the sequence of user flows and the assignment of laptops under constraints that ensured coverage across conditions. Average energy decreased from Run~1 (1453\,J) to Run~2 (905\,J) and then stabilized in Run~3 (836\,J); Figure~\ref{fig:run_order_effects} shows the corresponding pattern at the flow level. This indicates a brief learning effect but no evidence of fatigue or performance drift. Each flow--computer combination was executed by multiple participants (min.~4, max.~7), and between-participant variance in the mixed-effects analysis was small (ICC $=0.012$), suggesting that
idiosyncratic user behavior had limited influence on the results.

Repeated participation can affect execution efficiency and measured energy. We have mitigated order effects by randomizing both flow order and laptop assignment while ensuring coverage across conditions. Mean energy dropped from Run~1 (1453\,J) to Run~2 (905\,J) and then stabilized in Run~3 (836\,J) (Figure~\ref{fig:run_order_effects}), indicating a brief learning effect but no fatigue or drift. Each flow--computer condition included multiple participants (min.~4, max.~7), and participant-level variance was small (ICC $=0.012$), indicating that individual differences had little impact on the results.

\subsection{Construct validity: representing a service}
Each service was evaluated using one essential user flow. This is appropriate
for controlled model comparison, but service-level reporting may require
aggregating multiple essential flows to represent typical use.

\subsection{External validity: workload, hardware, and device scope}
The study evaluates browser-based, non-streaming services on four laptop
platforms spanning common Intel generations and OEM designs. Generalization
is limited in three respects: (i) the evaluated categories (shopping,
booking, navigation, news) represent only a subset of web interactions;
(ii) the dataset includes four laptop models, limiting vendor and
architectural diversity; and (iii) results are specific to laptops and may
not transfer to phones, tablets, or high-performance desktops with different
baseline power and scaling behavior. All interactions were executed on live
websites under real network conditions, supporting ecological validity for
laptop-based browsing within this scope.


%


\section{Discussion}

The results show that deviations between modeled and measured energy are not random but scale systematically with interaction length, revealing a systematic bias in the constant-power assumption of existing reporting models. The nearly perfect linear relation between Digst error and task duration confirms that this bias arises from the model formulation rather than from random measurement noise.



\subsection{Power Average of modern End-User Devices}
Our analysis of factors effect (Section \ref{sec:rq2}) started from the point of creating a new best-fit model based on the available data. This increased $R^2$ from a negative number, i.e., using the average measured energy usage is a better choice, to $94.6\%$. This shows that --when using an energy model with constant power assumption-- it in essential to calibrate the power average. Therefore, it is essential to compare our measured laptop power usage with that of the existing literature. 

The measured average laptop powers in our study (9--13~W) are substantially
lower than the fixed constants currently used in sustainability reporting
frameworks. For example, the Danish Agency for Digital Government (Digst)
assumes 22~W, and DIMPACT applies 15~W for browser-based scenarios.
Consequently, using inaccurate power constants can introduce duration-scaled
bias rather than random estimation noise.

Our observations align closely with those in Malmodin's recent study on long-term household measurements~\cite{Malmodin2023}, which report average powers between 6~W and 12~W with short-term fluctuations from 3~W to 25~W. In addition, Malmodin's results align well with the 10~W average derived in our best-fit model, which is realistic for browser-based workloads and representative of 
typical laptop energy behavior. 

In 2013, Schien et al.~\cite{Schien2013} observed average power dissipation of a laptop to be 32--37~W\footnote{ Supplementary material~\cite{Schien2013} report energy consumption for 10 minutes of reading text and video watching as 19.15~kJ and 22.01~kJ, respectively.}, and they found no statistically significant variation from idle
power when browsing text. This does not align with our results. For comparison, see Table~\ref{tab:power_pr_comp}, which lists the Energy Star measured idle power $P_{idle}$~\cite{hp_elitebook_830g6_2020,dell_latitude_e5250_2014,dell_latitude_e5550_2014,dell_latitude_7000series_2021}, the minimum observed power during execution, and the best-fit power average. Our best-fit power average is significantly higher than the Energy Star idle power and, in some cases, twice as high.

\begin{table}[b]
\centering
\caption{The laptops Energy Star $P_{idle}$, minimum observed power during execution, and the best-fit power  average.} 
\label{tab:power_pr_comp}
\begin{tabular}{c@{\,}lrrr} \toprule
\textbf{No.} & \textbf{Laptop} & $\mathbf{P}_{\mathbf{Idle}}$ & \textbf{Min. $P$} & \textbf{Best-fit $P$} \\ \midrule
1 & Dell Latitude 7490 & -- & 6.53\,W & 9.74\,W \\
2 & HP EliteBook 830 G6 & 4.88\,W & 7.30\,W & 9.56\,W \\
3 & Dell Latitude E5250 & 6.32\,W & 6.74\,W & 9.17\,W \\
4 & Dell Latitude E5550 & 5.12\,W & 10.92\,W & 12.43\,W \\ \bottomrule
\end{tabular}
\end{table}

Malmodin also noted that Apple’s M-series architectures achieve averages 
below 6~W, and subsequent silicon generations, such as the M4 and newer 
Intel hybrid designs, are expected to further reduce typical laptop power 
dissipation.

Overall, our measurements are consistent with ongoing
hardware efficiency improvements, implying that reporting models should be periodically recalibrated to keep assumed end-user power values representative of contemporary devices.
%
%
Alternatively, this suggests a need to explore alternative formulations that decouple energy estimation from pure duration, for instance by distinguishing a time-dependent baseline component from an activity-dependent component that captures variations in user interaction intensity.

\subsection{Improving Accuracy of Reporting Frameworks}
The purpose of reporting models is to enable fair and practically applicable comparisons of services.
Our analysis of factor effects (Section~\ref{sec:rq2}) started by estimating a new best-fit model from the available data. This increased $R^2$ from a negative value---indicating that the mean measured energy consumption would provide a better predictor---to $94.6\%$. Under the assumption that systematic error can be controlled, model precision becomes a central concern.

Across the evaluated alternatives, the model combining hardware characteristics with individual user flows achieved the highest explanatory power ($R^2=99.4\%$). In contrast, the scientific basis underlying the national models, as discussed by Preist et al., proposes averaging across Energy Star measurement results (i.e., across certified ``green'' devices) to reflect end-user device heterogeneity~\cite{Preist2019}. This emphasis on device diversity is compatible with the objective of practical reporting. However, operationalizing it in our setting would require measuring the same user flows on a sufficiently broad set of labelled devices to capture the plausible diversity of end-user hardware. Such a measurement effort may be infeasible, and methods for reducing this burden are beyond the scope of the present paper. Future work should therefore investigate the extent of device sampling required for robust reporting, and how this sampling can be achieved cost-effectively.

In some contexts, direct energy measurement may be impractical (e.g., for small services with limited user time). In our data, categories alone increased $R^2$ to $97.3\%$, suggesting that refining DIMPACT’s service categories could support model-based estimation when measurement is not feasible. 
While each category here included only two services, the results indicate that category-based refinement can improve model fit and support fairer comparisons among functionally similar services, akin to EU energy labelling.

\subsection{Video Streaming}
Although this study focuses on browser-based interactions, similar effects are expected in other end-user workloads, including streaming and video communication. In the \textit{TV2} flow, which included embedded autoplay video, we observed higher average power than on non-video pages, consistent with the controlled results of Herglotz et al.~\cite{Herglotz2025}.
 Whereas their study isolated and synchronized video playback under laboratory conditions, our setup captures realistic browsing sessions. Taken together, the results indicate that even brief embedded videos can dominate device-level energy use, motivating category-specific modeling in sustainability reporting.

\section{Conclusion \& Future Work}
This study demonstrates that the constant-power assumptions used in digital
sustainability reporting frameworks (e.g., Digst and DIMPACT) can
systematically overestimate end-user device energy use. Across realistic
browser-based interactions, we measured typical laptop powers of 9--13~W,
substantially below the 22~W and 15~W values assumed by these models.
Because energy is estimated as $P \times t$, any power mismatch induces a
duration-proportional bias, yielding systematic rather than random error.

Using a constant-power baseline, adding hardware and service-category parameters substantially improved model precision, showing that much of the variance reflects reproducible, context-dependent differences in devices and workloads. Thus, reporting can remain lightweight while becoming more accurate by adopting calibrated category-specific (and, where feasible, device-specific) power parameters.

Future work would benefit from pursuing three directions. First, expand the hardware base to a larger and more diverse set of contemporary laptops to improve the representativeness of derived parameters. Second, reduce systematic overestimation by moving beyond global constants and modeling energy as a function of task context, interaction type, and hardware characteristics. Third, evaluate the approach on additional software categories (e.g., streaming, productivity, and AI-assisted applications) and develop refined category definitions and standardized, fair user flows to support reproducible assessment.



\section*{Acknowledgments}
We thank participants of the \href{https://suits-25.github.io/}{Symposium on Sustainable IT Systems} for their contributions as participants and valuable insights, here among, especially Jens Malmodin, for discussions on the initial results. 
\bibliographystyle{IEEEtran}
\bibliography{references}

\end{document}